# Magnetometry of random AC magnetic fields using a single Nitrogen-Vacancy center

Abdelghani Laraoui[1], Jonathan S. Hodges[2], Carlos A. Meriles[1]

[1]*Department of Physics, City College of New York – CUNY, New York, NY 10031, USA*

[2]*Department of Electrical Engineering, Columbia University, New York, NY 10027, USA*

**Abstract**

We report on the use of a single NV center to probe fluctuating AC magnetic fields. Using engineered currents to induce random changes in the field amplitude and phase, we show that stochastic fluctuations reduce the NV center sensitivity and, in general, make the NV response field-dependent. We also introduce two modalities to determine the field spectral composition, unknown a priori in a practical application. One strategy capitalizes on the generation of AC-field-induced coherence 'revivals', while the other approach uses the time-tagged fluorescence intensity record from successive NV observations to reconstruct the AC field spectral density. These studies are relevant for magnetic sensing in scenarios where the field of interest has a non-trivial, stochastic behavior, such as sensing unpolarized nuclear spin ensembles at low static magnetic fields.

The proposal for using the spin of a Nitrogen-Vacancy (NV) center as a platform for nanoscale magnetic field sensing[1,2] has been followed by intense experimental and theoretical work.[1-7] Besides the ability to polarize and monitor single spins with high sensitivity, this system is appealing due to its long coherence times, exceeding 1 ms at room temperature, and the ability to localize its diamond host to nanometer scales. While initial proof-of-principle experiments have exploited this property to demonstrate detection with high sensitivity[3,4], the target magnetic field is either static or oscillates with a well-defined phase relative to the pulse sequence driving the NV spin (e.g., a gated field excitation). As discussed recently[8], these conditions do not necessarily reflect those found in a broad subset of applications in various biological and condensed matter systems, where the magnetic field fluctuates over time. Examples include the random fields stemming from stochastic charge transport[9] and the time-dependent 'spin noise' due to unpolarized electrons or nuclei adjacent to the sensor.[10,11] Here we focus on using a single NV as a probe for such stochastic fields and explore different detection strategies aimed at determining the field spectral composition.

The theory and practice of optically-detected magnetic resonance in single NV centers has been extensively described[12,13] and is only briefly summarized here. The NV pair has net spin $S=1$ and a triplet ground state with zero-field splitting of 2.87 GHz. Using a confocal microscope, it is possible to selectively address single centers; complete initialization of the ground state $m_s = 0$ takes place after ~1-3 μs illumination with green light. We use a small DC magnetic field, $B_0$, collinear with the crystal field to lift the degeneracy between the $m_s = \pm 1$ states, thus allowing one to selectively address a two-level system, e.g., between $m_s = 0$ and $m_s = -1$, by properly detuning the third level. The near field of a thin copper wire laid on the diamond crystal surface allows for spin manipulation; the fluorescence intensity resulting from a readout laser pulse at the end of the detection protocol provides the contrast (~30 %) necessary to discriminate the sensor spin state. In our experiments we use an antenna in the vicinity of the sample to generate an AC magnetic field of variable frequency, amplitude, and relative phase.

Fig. 1b shows the normalized NV fluorescence after the sequence in Fig. 1a: We observe the characteristic pattern of echo 'collapse and revival' from Larmor precession of neighboring $^{13}$C spins in the applied DC field.[14] With the application of an oscillatory, asynchronous field $B_A(t) = b_A \cos(2\pi \nu_A t + \varepsilon_A)$, evolution of the NV spin over the spin-echo time $\tau$ leads to an accumulated phase $\phi = \int_0^{\tau/2} \gamma_e B_A(t) dt - \int_{\tau/2}^{\tau} \gamma_e B_A(t) dt$, where $\gamma_e$ denotes the electron gyromagnetic ratio and the first (second) term refers to the phase accumulated during the first (second) half of the Hahn echo sequence. Figs. 1c and 1d show the NV response at the peak of the first revival as a function of the field intensity for the cases in which only the amplitude, $b_A$, or both the field amplitude and phase relative to the pulse sequence, $\varepsilon_A$, change randomly over time. These two regimes lead to remarkably different responses, the former resulting in long-lived oscillations and the latter in a rapid decay.

To interpret these observations we recall[11] that the normalized NV response to an external magnetic field is given by $S_{NV} = [(\alpha - \beta)/(\alpha + \beta)] \langle \cos\phi \rangle \exp(-\tau/T_{2e})^\delta$, where $\alpha$ ($\beta$) indicates the average number of photons per readout when the NV spin is in the $m_s=0$ ($m_s=-1$) states. $T_{2e}$ is the NV transverse relaxation time (240 μs), and $\delta$ is a constant (~1-3) characterizing the NV signal relaxation envelope. For a Hahn-echo sequence, $\langle\cos\phi\rangle = \int_0^{2\pi} d\varepsilon_A \int_0^\infty db_A p(\varepsilon_A, b_A)\cos\phi$, where $\phi = (2\gamma_e b_A/(\pi\nu_A))\times \sin^2(\pi\nu_A\tau/2)\sin(\pi\nu_A\tau+\varepsilon_A)$ is the accumulated phase and $d\varepsilon_A db_A p(\varepsilon_A, b_A)$ denotes the joint probability of an AC field with amplitude $b_A$ and relative phase $\varepsilon_A$. In our experiment, these variables change independently allowing us to write $p(\varepsilon_A, b_A) = p(\varepsilon_A)p(b_A) = p(b_A)/(2\pi)$, where we assumed uniform probability density for $\varepsilon_A$. Hence, we can rewrite the above formula as $\langle\cos\phi\rangle = \int_0^\infty db_A p(b_A) J_0(b_A/\tilde{b}_A)$, where $J_0$ denotes the zero-order Bessel function and $\tilde{b}_A = \pi\nu_A/(2\gamma_e \sin^2(\pi\nu_A\tau/2))$.

Solid lines in Figs. 1c and 1d reproduce the calculated response for the cases in which $p_\delta(b_A) = \delta(b_A - B_A^0)$ and $p_\Pi(b_A) = (1/B_A^0)\Pi(b_A/(2B_A^0))$, where $\delta$ is the 'delta' function and $\Pi$ denotes the 'uniform' distribution over the interval $[0, B_A^0]$. As inferred from the slopes in the curves of Figs. 1c and 1d, the system is most sensitive to changes in the root-mean-square (rms) field amplitude $b_A^{rms} \equiv \sqrt{\langle b_A^2\rangle}$ within the approximate range $[3\tilde{b}_A/2, 5\tilde{b}_A/2]$. An estimate of the sensitivity $\eta_{Async}^{AC}$ can be attained for $p_\delta(b_A)$. Using $b_A^{rms} = B_A^0$ we find $\langle\cos\phi\rangle \approx -(2\sqrt{2}/(\pi\sqrt{3}))(b_A^{rms}/\tilde{b}_A) - 3\pi/4)$ resulting in $\eta_{Async}^{AC} \approx 2\eta_{sync}^{AC}$, where $\eta_{Sync}^{AC} = (\pi/(2\gamma_e))\sqrt{\nu_A}\exp(1/(\nu_A T_{2e})^p)\sqrt{2(\alpha+\beta)}/(\alpha-\beta)$ denotes the 'synchronous' sensitivity and we assumed that the time separation $\tau_d$ between consecutive observations satisfies $\tau_d = \tau = 1/\nu_A$, a good approximation throughout our experiments. The above result represents a lower bound to the sensitivity in the case in which the probability density $p(b_A)$ has a finite width; for example, for $p(b_A) = p_\Pi(b_A)$ one obtains $\eta_{Async}^{AC} \approx 3\eta_{sync}^{AC}$.

We now focus on the regime $b_A^{rms} < \widetilde{b}_A$. Here we find $\langle \cos\phi \rangle \approx 1 - \left(b_A^{rms}/\left(2\widetilde{b}_A\right)\right)^3$, leading to $\eta_{Async}^{AC} \approx \left(\pi v_A/\left(\gamma_e b_A^{rms}\right)\right)\eta_{Sync}^{AC}$, where, as before, we assume $\tau_d = \tau = 1/v_A$. This field-dependent sensitivity reflects the 'quadratic' response of the magnetometer in the limit $b_A^{rms} \to 0$. Note that, unlike the typical approach in 'synchronous' experiments[3,4], this problem cannot be corrected by introducing a π/2-phase shift in the 'projection' pulse[11], or by supplementing the target field $B_A(t)$ with a known, in-phase field[3] $B_A^{Ref}(t) = b_A^{Ref} \cos(2\pi v_A t + \varepsilon_A)$. The former makes the NV response insensitive to external magnetic perturbations (to third or higher order), whereas the latter is inconsistent with the very assumption of random—i.e., unpredictable—phase fluctuations in the magnetic field source.

Extending the sensing time $\tau$ to an integer number $n$ of $B_A$ oscillations (i.e., $\tau \to \tau' = n/v_A$) provides partial remedy to this limitation, but in this case one has to include additional inversion pulses. The simplest solution is a Carr-Purcell train of $n$ units and the rescaling $\widetilde{b}_A \to \widetilde{b}_A' = \widetilde{b}_A/n$, which leads to $\eta_{Async}^{AC}{'} = \eta_{Async}^{AC} \exp\left((n-1)/v_A T_{2e}\right)/n^2$. The optimum number of inversions in the pulse train is thus $n_{Opt} = Round(2v_A T_{2e})$, i.e., the closest integer to $2v_A T_{2NV}$. This same strategy applied in the 'linear regime' $\left(b_A/\widetilde{b}_A\right) \in [1.5, 2.5]$ yields $\eta_{Async}^{AC}{'} = \eta_{Async}^{AC} \exp\left((n-1)/v_A T_{2e}\right)/n$ and $n_{Opt} = Round(v_A T_{2e})$.

As seen in Fig. 1d, one consequence of amplitude fluctuation is the partial loss of the spectral composition of the AC field. Fig. 2 shows a possible strategy that partly regains this information. Here we plot the NV response, $S_{NV}$, as a function of the sensing time $\tau$ for an asynchronous field of frequency $v_A$=12 kHz. To effectively eliminate the

$^{13}$C-induced beating in the resulting signal, the DC field amplitude—along with the corresponding microwave frequency—is swept so that the fluorescence collected at each time $\tau$ is that of the first echo revival. We engineered the AC field to exhibit either time-independent or randomly-fluctuating amplitudes (Figs. 2a and 2b, respectively). Similar to Fig. 1, the resulting signal is 'smoothened out' when the field amplitude changes over time. Nonetheless, the net phase accumulated by the NV spin is null when the AC field completes (at least) one full oscillation during each half of the Hahn-echo sequence, i.e., when $\tau=2/\nu_A$. Therefore, and in analogy with the $^{13}$C-induced pattern of 'collapses and revivals'[14], one can determine the main carrier frequency from the positions of the resulting train of maxima (of which Fig. 2 shows only the first). We measure $\tau_{max}\sim166$ μs, in reasonable agreement with the expected value $\tau_{calc}=1/(12\text{ kHz})=166.7$ μs.

An assumption in the example above is that the sensing time $\tau$ is shorter than the correlation time characterizing the source magnetic field. When this is the case, the two-time correlation function $K_{NV}(\Delta t) = \langle s_{NV}(t) s_{NV}(t+\Delta t) \rangle$—calculated from the record of $N$ successive observations $\{s_{NV}(t_1), s_{NV}(t_2)...s_{NV}(t_N)\}$—provides an alternate tool to characterize the spectral composition of the random field. More specifically, we have shown in prior theoretical work[11,10] that for single NV magnetometry the correlation signal is given by the formula $K_{NV}(\Delta t) = ((\alpha-\beta)/(\alpha+\beta))^2 \langle \sin\phi(t+\Delta t)\sin\phi(\Delta t) \rangle$. Via the Wiener-Khintchine theorem[15], we can get the underlying 'spectral density' from the Fourier transform $F_{NV}(\Delta\nu) = \int_{-\infty}^{\infty} K_{NV}(t) e^{-i2\pi\Delta\nu t} dt$. Note that because the Hahn-echo sequence has a central observation frequency $\nu_C=1/\tau$, $\Delta\nu$ must be interpreted as a shift relative to $\nu_C$. Similarly, the spectral bandwidth is defined by the inverse of the 'dwell time' $\tau_d$ separating two consecutive observations (here of order $\tau$, the duration of the Hahn-echo sequence).

Fig. 3 shows the measured spectral density for cases in which $b_A^0$ and $\varepsilon_A$ are engineered to fluctuate with a common, average correlation time of predefined duration. We observe maxima at $\Delta\nu_{max}\sim 5.2$ kHz, *twice* the difference $\Delta\nu_{AC}= \nu_A - \nu_C$ between the AC field ($\nu_A$=17 kHz) and observation frequencies ($\nu_A$=17 kHz). The latter mirrors the results of Fig. 2 (where the AC-field-induced 'revival' time doubles the period $1/\nu_A$), and derives from the bilinear nature of the correlation function. Progressively poorer signal-to-noise ratios are observed with decreasing correlation times, a manifestation of the difficulties faced when characterizing rapidly fluctuating, random fields. Note that processing the available information in the form of a power spectrum simplifies the identification of multiple spectral components; this is illustrated in the insert to Fig. 3 where the AC field was engineered to contain two close frequencies of similar amplitude.

These and similar methodologies could bode well for applications where the coherence time and spectral content of the field fluctuations reflect on the composition and/or dynamics of a material system under consideration. Examples include complex processes such as trans-membrane ion transfer, molecular diffusion in nanoporous systems, or spectral characterizations of heterogeneous nuclear spin species at low DC field.[8,10]

We thank G. Kowach for providing the diamond sample. We gratefully acknowledge L. Jiang, P. Cappellaro, J. Maze, C. Degen and M. Lukin for useful discussions. We acknowledge support from Research Corporation and from the National Science Foundation.

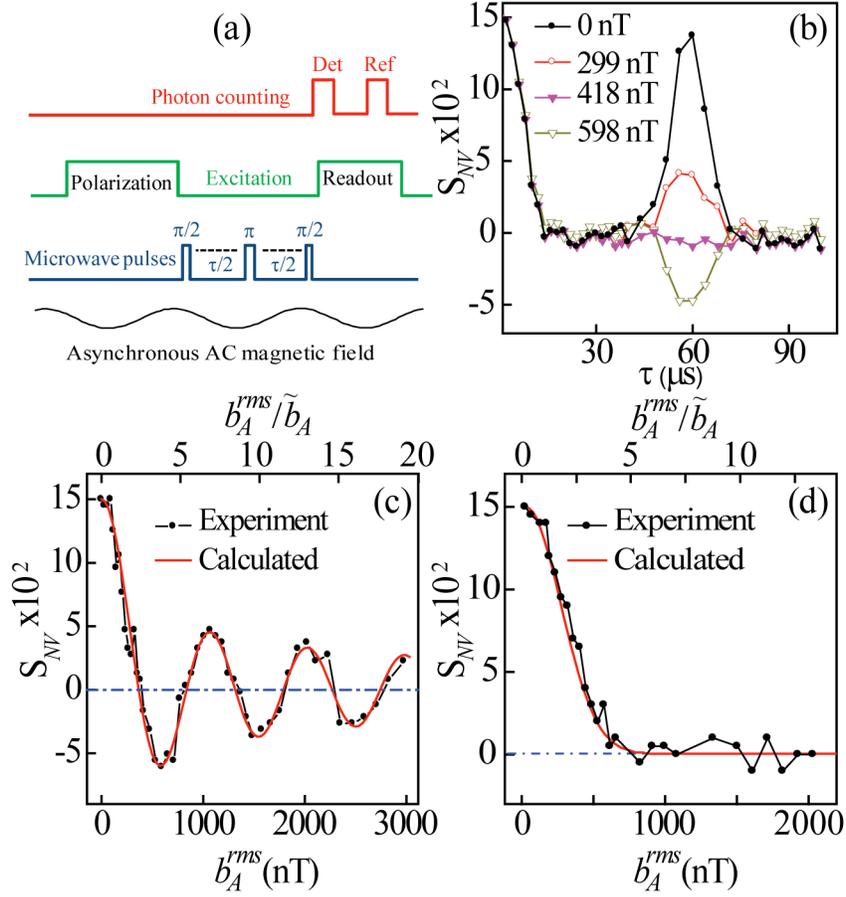

**Fig. 1: (a)** Basic detection protocol; the target AC field is assumed asynchronous relative to the microwave pulse sequence. **(b)** Spin echo signal as a function of the total evolution time τ for AC fields of various amplitudes. **(c)** Average echo amplitude at the time of the first revival (60 μs) as a function of the rms field amplitude for the case $p(b_A) = \delta(b_A - B_A^0)$ **(d)** Same as in (c) but for the case $p(b_A) = (1/B_A^0)\Pi(b_A/(2B_A^0))$. In all experiments, the AC field has a fixed frequency of 17 kHz; the DC field, of amplitude 3.1 mT, is aligned along the crystal field. Each point represents the average of $5 \times 10^5$ repeats.

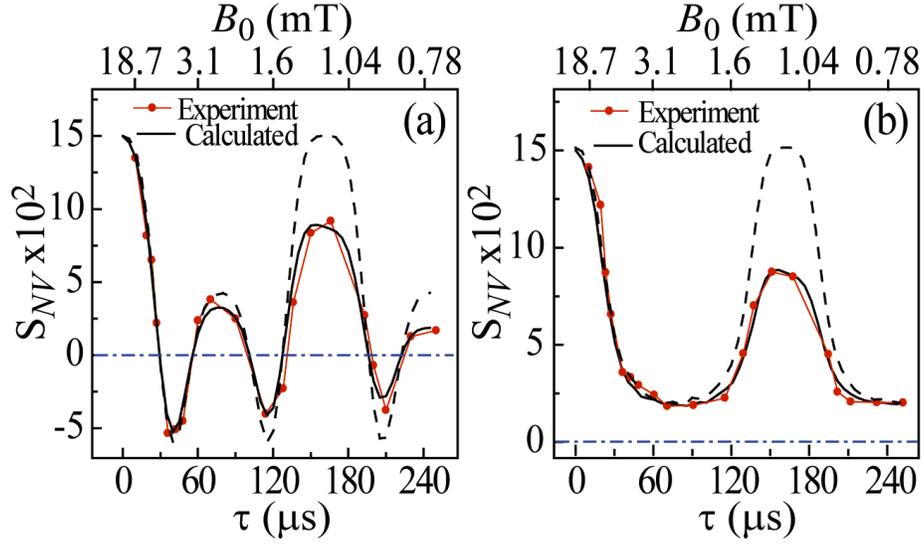

**Fig. 2: (a)** NV center fractional fluorescence change as a function of the sensing time $\tau$ in the presence of an asynchronous AC field of carrier frequency $\nu_A$=12 kHz and fixed amplitude $B_A^0$=714 nT. For reference, the upper horizontal axis indicates the value of the DC field at each time $\tau$. **(b)** Same as in (a) but for the case in which the AC field amplitude fluctuates randomly within the range 0-714 nT. In (a) and (b) the dashed curve reproduces the calculated signal dependence in the absence of NV decoherence. Each experimental point represents the average of $5\times10^5$ repeats.

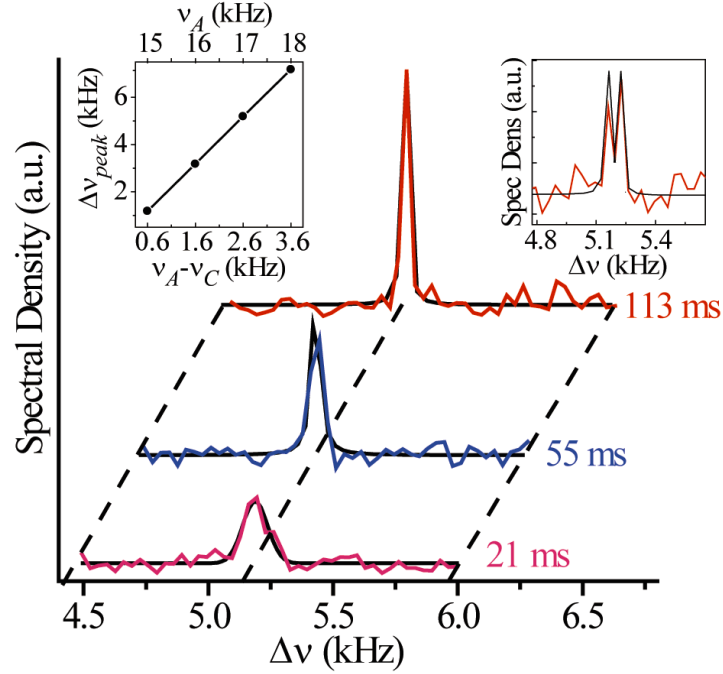

**Fig. 3: (Main)** Fourier transform of the NV center correlation function $K_{NV}$ as determined from the record of $2 \times 10^8$ consecutive observations in the presence of an asynchronous AC field of increasingly shorter correlation time. Each data entry is the number of photons collected after a Hahn-echo sequence ($\tau$=60 μs) in the presence of a $B_0$ field of 3.1 mT. The time separation between consecutive observations is $\tau_d$=70 μs; the AC field amplitude is 418 nT and $v_A$=17 kHz. (Left insert) Measured shift $\Delta v_{peak}$ corresponding to the position of the spectrum maximum for different input frequencies $v_A$; the slope in a linear fit is 2.0. (Right insert) Spectral density in the case in which the input AC field has two, close-by frequencies.

11